\documentclass{eptcs}
\providecommand{\event}{GANDALF 2015} 

\usepackage{microtype}
\usepackage{amsmath,amssymb,stmaryrd,xspace,tikz,wasysym}
\usepackage{algorithm, algorithmicx}
\usepackage{algpseudocode}
\usepackage{pgfplots}
\usepackage{pgfplotstable}
\usepackage{thm-restate}

\usetikzlibrary{arrows,automata,shapes,calc,fit}

\newtheorem{defi}{Definition}
\newtheorem{theo}{Theorem}
\newtheorem{prop}{Proposition}
\newtheorem{lemm}{Lemma}
\newtheorem{coro}{Corollary}
\newtheorem{exam}{Example}

\newenvironment{definition}{\begin{defi} \rm }{\end{defi}}
\newenvironment{theorem}{\begin{theo} \rm }{\end{theo}}
\newenvironment{proposition}{\begin{prop} \rm }{\end{prop}}

\newenvironment{example}{\begin{exam} \rm }{\end{exam}}

\def\nat{\mathbb{N}}

\newcommand{\plays}[2]{\Pi(#1,#2)}

\renewcommand{\i}{\ensuremath{\ocircle}\xspace}
\newcommand{\odd}{\ensuremath{\square}\xspace}
\newcommand{\even}{\ensuremath{\Diamond}\xspace}
\newcommand{\player}{\ensuremath{\ocircle}\xspace}

\newcommand{\post}[1]{\ensuremath{#1^{\bullet}}}

\usepackage{todonotes}

\newcommand{\pnot}[1]{\bar{#1}}

\newcommand{\attrsym}{\ensuremath{\textit{Attr}}}
\newcommand{\attr}[3][]{\ensuremath{{#2}{\text{-}}\attrsym^{#1}(#3)}}
\newcommand{\myattr}[3]{\ensuremath{{#2}{\text{-}}\attrsym^{#1}(#3)}}

\newcommand{\attrW}[3]{\ensuremath{\mathop{{#1}{\text{-}}\attrsym_{#3}(#2)}}}
\newcommand{\myattrW}[4]{\ensuremath{\mathop{{#2}{\text{-}}\attrsym^{#1}_{#4}(#3)}}}

\newcommand{\winsubodd}[1]{\textsf{Win}_{\odd}(#1)}
\newcommand{\winsubeven}[1]{\textsf{Win}_{\even}(#1)}
\newcommand{\minprio}[1]{\textsf{min}_{\priosym}(#1)}

\newcommand{\ie}{\emph{i.e.}\xspace}
\newcommand{\eg}{\emph{e.g.}\xspace}
\newcommand{\viz}{\emph{viz.}\xspace}

\newcommand{\oftype}{{:}}

\newcommand{\priosym}{\mathcal{P}}
\newcommand{\prio}[1]{\priosym(#1)}

\newcommand{\strategy}[1]{\mathbb{S}_{#1}}
\newcommand{\memstrategy}[1]{\mathbb{S}_{#1}^*}

\title{Improvement in Small Progress Measures}
\def\titlerunning{Improvement in Small Progress Measures}
\author{Maciej Gazda and Tim A.C. Willemse
\institute{Eindhoven University of Technology, The Netherlands}
}
\def\authorrunning{M. Gazda and T.A.C. Willemse}

\def\runtimeceil{O(dm \cdot (n/\lceil d / 2 \rceil)^{\lceil d/2 \rceil})}
\def\runtimefloor{O(dm \cdot (n/\lfloor d / 2 \rfloor)^{\lfloor d/2 \rfloor})}
\def\floord2{{\lfloor d/2 \rfloor}}
\def\ceild2{{\lceil d/2 \rceil}}

\begin{document}
\maketitle

\begin{abstract}
Small Progress Measures is one of the classical parity game solving
algorithms. For games with $n$ vertices, $m$ edges and $d$ different
priorities, the original algorithm computes the winning regions and
a winning strategy for one of the players in $\runtimefloor$ time.
Computing a winning strategy for the other player requires a re-run
of the algorithm on that player's winning region, thus increasing
the runtime complexity to $\runtimeceil$ for computing the winning
regions and winning strategies for both players. We modify the
algorithm so that it derives the winning strategy for both players
in one pass. This reduces the upper bound on strategy derivation
for SPM to $\runtimefloor$. At the basis of our modification is a
novel operational interpretation of the least progress measure that
we provide.  \end{abstract}

\section{Introduction}

A parity game \cite{EJ:91,McN:93,Zie:98} is an infinite duration
game played on a directed graph by two players called \emph{even}
and \emph{odd}. Each vertex in the graph is owned by one of the
players, and labelled with a natural number, called a priority.
The game is played by pushing a token along the edges in the graph;
the choice where to move next is made by the owner of the vertex
on which the token currently resides.  The winner of the thus
constructed play is determined by the parity of the minimal (or
maximal, depending on the convention) priority that occurs infinitely
often, and the winner of a vertex is the player who has a \emph{strategy}
to force every play originating from that vertex to be winning for
her.  Parity games are positionally determined; that is, each vertex is won by
some player~\cite{McN:93}, and  each player has a positional winning strategy
on her winning region. \emph{Solving} a game essentially means
deciding which player wins which vertices in the game.

Parity games play an important role in several foundational results;
for instance, they allow for an elegant simplification of the hard
part of Rabin's proof of the decidability of a monadic second-order
theory, and a number of decision problems of importance can be
reduced to deciding the winner in parity games. For instance, the
model checking problem for the modal $\mu$-calculus is equivalent,
via a polynomial-time reduction, to the problem of solving parity
games~\cite{EJS:93,SS:98}; this is of importance in computer-aided
verification.  Winning strategies for the players play a crucial
role in supervisory control of discrete event systems, in which
such strategies are instrumental in constructing a supervisor that
controls a plant such that it reaches its control objectives and
avoids bad situations; see \eg~\cite{AVW:03} and the references
therein. In model checking, winning strategies are essential in
reporting witnesses and counterexamples, see~\cite{SS:98}.\medskip

A major impetus driving research in parity games is their computational
status: the solution problem lies in UP and coUP and is, despite
the continued research effort, not known to be in PTIME.  
Deterministic algorithms for solving parity games include the
classical \emph{recursive algorithm}~\cite{Zie:98} and
\emph{small progress measures} (SPM) algorithm~\cite{Jur:00}, the
\emph{bigstep} algorithm~\cite{Sch:07}, the deterministic subexponential
algorithm~\cite{JPZ:06} and a class of strategy improvement
algorithms~\cite{VJ:00,Sch:08,Fea:10}.

Strategy improvement algorithms were long perceived as likely
candidates for solving parity games in polynomial time, but, save
a recently introduced symmetric variant~\cite{STV:15}, they were
ultimately proven to be exponential in the worst-case~\cite{Fri:11}.
The deterministic subexponential algorithm, running in $n^{O(\sqrt{n})}$
where $n$ is the number of vertices in the game, is a modification
of the recursive algorithm which
itself runs in $O(m \cdot n^d)$, where $m$ is the number of edges
and $d$ is the number of different priorities in the game. Bigstep,
which runs in $O(m \cdot (\kappa n /d)^{\gamma(d)})$, where $\kappa$
is a small constant and $\gamma(d) \approx d/3$, is a combination
of the recursive algorithm and 
the SPM algorithm.
This latter algorithm solves a game in $\runtimefloor$.

Somewhat surprisingly, our knowledge of classical algorithms such
as SPM and the recursive algorithm is still far from complete. For
instance, the recursive algorithm is regarded as one of the best
algorithms in practice, which is corroborated by experiments~\cite{FL:09}.
However, until our recent work~\cite{GW:13} where we showed the
algorithm is well-behaved on several important classes of parity
games, there was no satisfactory explanation why this would be the
case. In a similar vein, in \emph{ibid.} we provided tighter bounds
on the worst-case running time, but so far, no tight bounds for
this seemingly simple algorithm have been established.  We expect
that if improvements on the upper bound on the parity game solving
problem can be made, such improvements will come from improvements
in, or through a better understanding of the classical algorithms;
this expectation is fuelled by the fact that these classical
algorithms and the ideas behind them have been at the basis of the
currently optimal algorithms.\medskip

In this paper, we focus on Jurdzi\'nski's
small progress measures algorithm. Using a fixpoint computation,
it computes a \emph{progress measure}, a labelling of vertices,
that witnesses the existence of winning strategies.   In general, no clear,
intuitive interpretation of the information contained in the progress
measures has been given, and the mechanics of the algorithm remain rather
technical.  This is in contrast to the self-explanatory
recursive algorithm, and the strategy improvement algorithm, where,
thanks to the ordering of plays according to profiles, at every step,
one has a clear picture of the currently known best strategy. Apart
from Jurdzi{\'n}ski's original article, some additional insight was
offered by Klauck and Schewe.  In~\cite{2001automata},
Klauck defines a specific parity progress measure for a solitaire game with
only even simple cycles and explains that it has a particular interpretation,
but his parity progress measure is not generally related to the measure
computed by the SPM algorithm (not even in the setting of solitaire games).
Schewe, in his paper on
\emph{bigstep} \cite{Sch:07}, analyses progress measures with restricted
codomains and shows how they can be utilised to efficiently detect small dominions.
Our \emph{first} contribution  is to provide a better understanding of
these progress measures and the intermediate values in the fixpoint
computation, see Section~\ref{sec:interpretation}.  By doing so,
a better understanding of the algorithm itself is obtained.

Progress measures come in two flavours, \viz even-and odd-biased,
and their computation time is bounded by either $\runtimefloor$ or
$\runtimeceil$, depending on the parity of the extreme priorities that
occur in the game.
From an even-biased progress measure, one immediately obtains winning
regions for \emph{both} players, but only a winning strategy for
player even on its winning region and not for her opponent. Likewise
for odd-biased progress measures. Obtaining the winning strategy
for an opponent thus requires re-running the algorithm on the
opponent's winning region.  Note that the effort that needs to be
taken to obtain a strategy in the same time bound as the winning
region stems from a more general feature of parity games: a winning
partition in itself does not allow one to efficiently compute a
winning strategy (unless there is an efficient algorithm for solving
parity games). This basic result, which we nevertheless were unable
to find in the literature, is formalised in Section \ref{sec:strategy}.

An essential consequence of this is that the algorithm solves a
parity game in $\runtimefloor$, as one can always compute one of
the two types of progress measures in the shorter time bound. Contrary
to what is stated in~\cite{Jur:00},
the same reasoning does not apply to computing the winning strategy for
a fixed player; this still requires $\runtimeceil$ in the worst
case,  as also observed by Schewe in~\cite{Sch:07}.
An intriguing open problem is
whether it is at all possible to derive the winning strategies for
both players while computing one type of measure only, as this would
lower the exponent in the time bound to $\floord2$.  Our \emph{second}
contribution is to
give an affirmative answer to the above problem.  We modify the
generic SPM by picking a partial strategy when a vertex won by
player $\odd$ is discovered, and subsequently adjust the lifting
policy so that it prioritises the area which contains an $\odd$-dominion.
Both steps are inspired by the interpretation of the progress measures
that we discuss in Section~\ref{sec:interpretation}.
Like the original algorithm, our solution, which we present in
Section~\ref{sec:strategy}, still works in polynomial
space. 
Formal proofs of all results can be found in our technical report~\cite{GW:14}.
\section{Parity Games}

\label{sec:parity_games}

A parity game is an infinite duration game, played by players \emph{odd},
denoted by $\odd$ and \emph{even}, denoted by $\even$, on a directed,
finite graph. The game is formally defined as follows.

\begin{figure}[h!]
\centering
\begin{tikzpicture}[>=stealth']
\tikzstyle{every node}=[draw, inner sep=2pt, outer sep=0pt, node distance=40pt];
  \node[label=above:{\scriptsize $v_6$},shape=rectangle,minimum size=15] (y4)                    {\scriptsize 1};
  \node[label=above:{\scriptsize $v_5$},shape=rectangle,minimum size=15] (y5) [left of=y4,xshift=-20pt]                    {\scriptsize 0};
  \node[label=above:{\scriptsize $v_4$},shape=rectangle,minimum size=15]   (y6) [left of=y5,xshift=-20pt]                   {\scriptsize 2};
  \node[label=above:{\scriptsize $v_3$},shape=diamond,minimum size=19] (y7) [left of=y6,xshift=-20pt]                   {\scriptsize 3};
  \node[label=above:{\scriptsize $v_2$},shape=diamond,minimum size=19]   (y8) [left of=y7,xshift=-20pt]                   {\scriptsize 3};
  \node[label=above:{\scriptsize $v_1$},shape=rectangle,minimum size=15] (y9) [left of=y8,xshift=-20pt]                   {\scriptsize 0};

\path[->]
  (y4.south) edge[bend left] (y6.south)
  (y5) edge (y4)
  (y6) edge (y5)
  (y6.north) edge[bend left] (y4.north) 
  (y7) edge (y6) edge[bend right] (y8) 
  (y8) edge[bend right] (y7) edge[bend right] (y9)
  (y9) edge[loop left] (y9) edge[bend right] (y8) 
;

\end{tikzpicture}
\caption{A simple parity game with 4 different priorities, in which 4 vertices are 
owned by player odd and 2 vertices are owned by player even.}
\label{fig:example}
\end{figure}
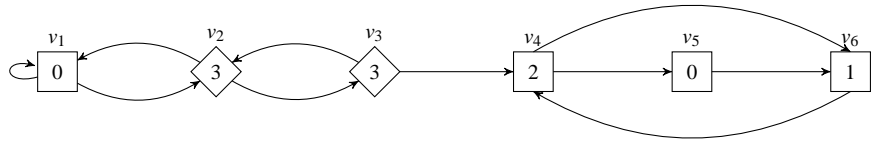

\begin{definition}
A parity game is a tuple $(V, E, \priosym, (V_\even,V_\odd))$, where
\begin{itemize}
\item $V$ is a finite set of vertices, partitioned in a set $V_\even$ of
vertices owned by player $\even$, and a set of vertices $V_\odd$ 
owned by player $\odd$,
\item $E \subseteq V \times V$ is a total edge relation, \ie all vertices
have at least one outgoing edge,
\item $\priosym \oftype V \to \nat$ is a priority function that assigns
priorities to vertices.
\end{itemize}
\end{definition}
Henceforth, we assume that $\i$ denotes an arbitrary
player; that is $\i \in \{\odd,\even\}$. We write $\pnot{\i}$ for
$\i$'s opponent:  $\pnot{\even}=\odd$ and $\pnot{\odd}=\even$.
Parity games are depicted as graphs; diamond-shaped nodes
represent vertices owned by player $\even$, box-shaped nodes
represent vertices owned by player $\odd$ and the priority assigned
to a vertex is written inside the node, see the game depicted in
Figure~\ref{fig:example}.

Throughout this section, assume that $G = (V,E,\priosym, (V_\even,
V_\odd))$ is an arbitrary parity game.  We write $v \to w$ iff
$(v,w) \in E$, and we write $\post{v}$ to denote the set of successors of
$v$, \ie\ $\{w \in V ~|~ v \to w\}$. For a set of vertices $W \subseteq V$, we will denote
the minimal priority occurring in $W$ with $\minprio{W}$; by 
$V_i$ we denote the set of vertices with priority $i$; likewise
for $V_{\le k}$.   For
a set $A \subseteq V$, we define $G \cap A$ as the structure $(A,
E \cap (A \times A), \priosym|_A, (V_\even \cap A, V_\odd \cap A))$;
the structure $G \setminus A$ is defined as $G \cap (V \setminus
A)$. The structures $G \cap A$ and $G \setminus A$ are again 
parity games if their edge relations are total. 

\paragraph*{Plays and Strategies}
A sequence of vertices $v_1, \ldots, v_n$ is a \emph{path} if $v_m \to
v_{m+1}$ for all $1 \leq m < n$.  Infinite paths are defined in a similar
manner. We write $\pi_i$ to denote the $i^\textrm{th}$ vertex in a path $\pi$.

A game starts by placing a token on some vertex $v \in V$.  Players
$\even$ and $\odd$ move the token indefinitely according to a single
simple rule: if the token is on some vertex that belongs to player
$\i$, that player moves the token to an adjacent vertex.
An infinite sequence of vertices created this way is called a
\emph{play}.  The \emph{parity} of the \emph{least} priority that occurs
infinitely often on a play defines the \emph{winner} of the play:
player $\even$ wins if, and only if this priority is even. 

A \emph{strategy} for player $\i$ is a partial function $\sigma
\oftype V^+ \to V$ satisfying that whenever it is defined for a
finite path $u_1 \cdots u_n \in V^+$, both $u_n \in V_{\i}$ and
$\sigma(u_1\cdots u_n) \in \post{u_n}$. We denote the set of
strategies of player $\player$ by $\memstrategy{\player}$.  
An infinite play $u_1\
u_2\ u_3 \cdots$ is \emph{consistent} with a strategy $\sigma$ if
for all prefixes $u_1 \cdots u_n$ of the play for which $\sigma(u_1\
\cdots u_n)$ is defined, $u_{n+1} = \sigma(u_1 \cdots u_n)$. The
set of all plays, consistent with some strategy $\sigma$, and starting
in $v$ is denoted $\plays{\sigma}{v}$. Some
strategy $\sigma$ is winning for player $\i$ from vertex $v$ iff
all plays consistent with $\sigma$ are won by player $\i$.  Vertex
$v$ is won by player $\i$ whenever she has a winning strategy for
all plays starting in vertex $v$.  Parity games are
\emph{determined}~\cite{EJ:91}, meaning that every vertex is won
by one of the players; they are even \emph{positionally determined},
meaning that if $\i$ wins a vertex then she has a winning
\emph{positional strategy}: a strategy that determines where to
move the token next based solely on the vertex on which the token
currently resides. Such strategies can be represented by a function
$\sigma\oftype V_{\i} \to V$, and the set of all such strategies for
player $\player$ is denoted $\strategy{\player}$.  \emph{Solving} a parity game $G$
essentially means computing the partition $(\winsubeven{G},\winsubodd{G})$
of $V$ into vertices won by player $\even$ and player $\odd$,
respectively.

\begin{example}
In the parity game depicted in Figure~\ref{fig:example}, vertices
$v_1, v_2$ and $v_3$ are won by player $\even$ while vertices $v_4, v_5$ and $v_6$ are
won by player $\odd$. A winning positional strategy for player $\even$
is to play from $v_2$ to $v_1$ and from $v_3$ to $v_2$. A winning
strategy for $\odd$ is to move from $v_4$ to $v_6$ and
from $v_5$ to $v_6$.
\end{example}

\paragraph*{Attractors and Dominions}
An $\i$-\emph{attractor} into a
set $U \subseteq V$ contains all those vertices from which player $\i$ can
force any play to $U$; it is formally defined as follows.
\begin{definition} The $\i$-\emph{attractor} into a set $U \subseteq V$,
denoted $\attr{\i}{U}$, is the least set $A \subseteq V$ satisfying:
\begin{enumerate}
 \item $U \subseteq A$
 \item
 \begin{enumerate}
 \item if $w \in V_{\i}$ and $\post{w} \cap A \neq \emptyset$, then $w \in A$
 \item if $w \in V_{\pnot{\i}}$ and $\post{w} \subseteq A$, then $w \in A$
 \end{enumerate}
\end{enumerate}

\end{definition}
Observe that the complement of an attractor set of any subset of a
parity game induces a parity game, \ie $G  \setminus \attr{\i}{U}$
for any $U$ and $\i$ is a well-defined parity game.  Whenever we
refer to an \emph{attractor strategy} associated with $\attr{\i}{U}$,
we mean the positional strategy that player $\i$ can employ to force
play to $U$; such a strategy can be computed in $\mathcal{O}(|V|+|E|)$
using a straightforward fixpoint iteration.

A set of vertices $U$ is an $\i$-dominion whenever there is a
strategy $\sigma$ for player $\i$ such that every play starting in
$U$ and conforming to $\sigma$ is winning for $\i$ and stays within
$U$. A game is a \emph{paradise} for player $\i$ if the entire game
is an $\i$-dominion.

\begin{example}
Reconsider the parity game of Figure~\ref{fig:example}. We have
$\attr{\even}{v_2} = \{v_2,v_3\}$ and $\attr{\odd}{v_4} = \{v_4,v_5,v_6\}$.
The winning region $\{v_1,v_2,v_3\}$ is an $\even$-dominion, but the
subset $\{v_2,v_3\}$ is not; the set $\{v_4,v_6\}$ is an $\odd$-dominion.
\end{example}

\newcommand{\Nat}{\ensuremath{\mathbb{N}}}
\newcommand{\prog}[3]{\ensuremath{\textsf{Prog}(#1,#2,#3)}}
\newcommand{\lift}[2]{\ensuremath{\textsf{Lift}(#1,#2)}}
\newcommand{\liftodd}[2]{\ensuremath{\textsf{Lift}_\odd(#1,#2)}}
\def\progname{\textsf{Prog}}
\def\liftname{\textsf{Lift}}

\section{Jurdzi\'nski's Small Progress Measures Algorithm}
\label{sec:spm}

The SPM algorithm works by computing a \emph{measure} associated with
each vertex. This measure is such
that it decreases along the play with each ``bad'' odd priority
encountered, and only increases upon reaching a beneficial even
priority. In what follows, we recapitulate the essentials for defining
and studying the SPM algorithm; our presentation is---as in
the original paper by Jurdzi\'nski---from the perspective of player
$\even$. \medskip

Let $G = (V, E, \priosym, (V_{\even}, V_{\odd}))$ be a parity game.
Let $d$ be a positive number and let $\alpha
\in \Nat^d$ be a \emph{$d$-tuple} of natural numbers.  We number
its components from $0$ to $d-1$, \ie $\alpha = (\alpha_0, \alpha_1,
\dots, \alpha_{d-1})$, and let $<$ on such $d$-tuples be given by
the lexicographic ordering. We define a derived ordering $<_i$ on $k$-tuples and
$l$-tuples of natural numbers as follows:
\[
(\alpha_0, \alpha_1, \dots, \alpha_k) <_i (\beta_0,\beta_1,\dots,\beta_l)
\text{ iff }
(\alpha_0, \alpha_1, \dots, \alpha_i) < (\beta_0,\beta_1,\dots,\beta_i)
\]
where, if $i > k$ or $i >l$, the tuples are suffixed with $0$s.
Analogously, we write $\alpha =_i \beta$ to denote that $\alpha$ and
$\beta$ are identical up-to and including position $i$.
The derived ordering provides enough information to
reason about the interesting bits of plays: when encountering a
priority $i$ in a play, we are only interested in how often we can
or must visit vertices of a more significant (\ie smaller) priority
than $i$,  whereas we no longer care about the less significant priorities that we shall encounter; a more precise interpretation of the information that will be encoded will be given in Section~\ref{sec:interpretation}. 
\medskip

Now, assume from hereon that $d-1$ is the largest priority occurring in
$G$; \ie, $d$ is one larger than the largest priority
in $G$.  For $i \in \Nat$, let $n_i = |V_i|$. 
Define $\mathbb{M}^{\even} \subseteq \Nat^d \cup \{\top\}$, as the largest
set containing $\top$
($\top \notin \Nat^d$) and
only those $d$-tuples with $0$ on \emph{even} positions and
natural numbers $\leq$ $n_i$ on \emph{odd} positions $i$.

The lexicographical ordering $<$ and the family of orderings
$<_i$ on $d$-tuples is extended to an ordering on $\mathbb{M}^\even$
by setting $\alpha < \top$ and $\top = \top$. 
Let $\rho {:} V \to \mathbb{M}^\even$ and suppose $w \in \post{v}$. Then
$\prog{\rho}{v}{w}$ is the least $m \in \mathbb{M}^{\even}$, such that
\begin{itemize}
  \item $m \geq_{\prio{v}} \rho(w)$ if $\prio{v}$ is even,
  \item $m >_{\prio{v}} \rho(w)$, or $m = \rho(w) = \top$ if $\prio{v}$ is odd.
\end{itemize}

\begin{definition} Function
$\rho$ is a \emph{game parity progress measure}
if for all $v \in V$:
\begin{itemize}
  \item if $v \in V_{\even}$, then for some $w \in \post{v}$, $\rho(v) \geq_{\prio{v}} \prog{\rho}{v}{w}$
  \item if $v \in V_{\odd}$, then for all $w \in \post{v}$, $\rho(v) \geq_{\prio{v}}\prog{\rho}{v}{w}$
\end{itemize}
\end{definition}
The following proposition is due to Jurdzi\'nski~\cite{Jur:00}; it shows that
the least game parity progress measure characterises the winning regions of a
parity game.

\begin{proposition}\label{prop:jurdzinski}
If $\rho$ is the \emph{least} game parity progress measure for $G$, 
then for all $v \in V$:
$\rho(v) \neq \top$ iff $v \in \winsubeven{G}$.
\end{proposition}
The least game parity progress measure can be described as the least fixpoint
of a monotone transformer on the complete lattice we define next. Let
$\rho,\rho' {:} V \to \mathbb{M}^\even$ and define
$\rho \sqsubseteq \rho'$ as $\rho(v) \leq \rho'(v)$ for all $v \in V$. We write
$\rho \sqsubset \rho'$ if $\rho \sqsubseteq \rho'$ and $\rho \not= \rho'$. Then
the set of all functions $V \to \mathbb{M}^\even$ with $\sqsubseteq$ is a complete
lattice. The monotone transformer defined on this set is given as follows:
\[
\lift{\rho}{v} = \begin{cases}
\rho[v \mapsto \max \{ \rho(v), \min\{ \prog{\rho}{v}{w} \mid v \to w \} \}] &
\text{if $v \in V_{\even}$}\\
\rho[v \mapsto \max \{ \rho(v), \max\{ \prog{\rho}{v}{w} \mid v \to w \} \}] &
\text{if $v \in V_{\odd}$}
\end{cases}
\]
As a consequence of Tarski's fixpoint theorem, we know the least fixpoint of the
above monotone transformer exists and can be computed using a standard fixpoint iteration
scheme. This leads to the original SPM algorithm, see Algorithm~\ref{alg:spm}.
\begin{algorithm}[h!t]
\begin{algorithmic}[1]
\Function{SPM}{$G$}
\State \emph{\textbf{Input} $G = (V, E, \priosym, (V_\even, V_\odd))$}
\State \emph{\textbf{Output} Winning partition $(\winsubeven{G},\winsubodd{G})$}
\State $\rho  \gets \lambda v \in V.~(0, \dots, 0)$
\While{$\rho \sqsubset \lift{\rho}{v}$ for some $v \in V$}
  \State $\rho \gets \lift{\rho}{v}$ for some $v \in V$ such that $\rho \sqsubset \lift{\rho}{v}$
\EndWhile
\State \Return $(\{v \in V ~|~ \rho(v) \not= \top\}, \{v \in V ~|~ \rho(v) = \top\})$
\EndFunction
\end{algorithmic}
\caption{The original Small Progress Measures Algorithm}
\label{alg:spm}
\end{algorithm}
Upon termination of the iteration within the SPM algorithm, the
computed game parity progress measure $\rho$ is used to compute the
sets $\winsubeven{G}$ and $\winsubodd{G}$ of vertices won by player $\even$ and
$\odd$, respectively.
\begin{theorem} Algorithm~\ref{alg:spm} solves a parity
game in $\runtimefloor$, see~\cite{Jur:00}.
\end{theorem}

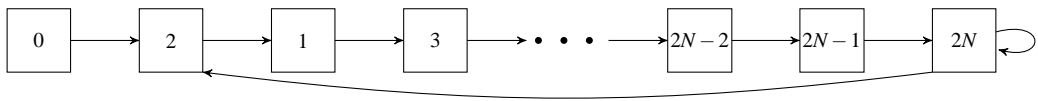
\begin{figure}[h]
\centering
\begin{tikzpicture}[>=stealth']
\tikzstyle{every node}=[draw, inner sep=1pt, outer sep=0pt, node distance=30pt];
  \node[shape=rectangle,minimum size=24] (y0)                            {\scriptsize $0$};
  \node[shape=rectangle,minimum size=24] (y1) [right of=y0,xshift=20pt]  {\scriptsize $2$};
  \node[shape=rectangle,minimum size=24] (y2) [right of=y1,xshift=20pt]  {\scriptsize $1$};
  \node[shape=rectangle,minimum size=24] (y21) [right of=y2,xshift=20pt]  {\scriptsize $3$};
  \node[draw=none] (y3) [right of=y21,xshift=20pt] {\Huge $\dots$};
  \node[shape=rectangle,minimum size=24] (y4) [right of=y3,xshift=20pt]  {\scriptsize $2N-2$};
  \node[shape=rectangle,minimum size=24] (y5) [right of=y4,xshift=20pt]  {\scriptsize $2N-1$};
  \node[shape=rectangle,minimum size=24] (y6) [right of=y5,xshift=20pt]  {\scriptsize $2N$};

\path[->]
  (y0) edge (y1)
  (y1) edge (y2)
  (y2) edge (y21)
  (y21) edge (y3)
  (y3) edge (y4)
  (y4) edge (y5)
  (y5) edge (y6)
  (y6.south west) edge[bend left=7] (y1.south east) 
  (y6) edge[loop right] (y6)
;
\end{tikzpicture}
\caption{A parity game won by player $\odd$. Solving the game using
$\mathbb{M}^{\even}$, the first $\top$ value is reached after the first full pass
of the cycle containing priority $1$ ($O(N^2)$ using a fair lifting
strategy), and it will then propagate through the game.
Solving the dual game, or using $\mathbb{M}^{\odd}$ takes exponential
time to lift the node with priority $2N$.}

\label{fig:tops_faster}
\end{figure}

The runtime complexity of SPM is obtained by considering the more optimal
runtime of solving a game $G$, or $G$'s `dual'; the latter is obtained by shifting
all priorities by one and swapping ownership of all vertices (alternatively,
a `dual' algorithm can be used, computing with a domain $\mathbb{M}^\odd$). 
The runtime complexity for computing winning strategies for both players using
the SPM algorithm is worse than the runtime complexity of solving the game.  
A winning strategy
$\sigma_\even {:} V_\even \to V$ for player $\even$ can be extracted
from $\rho$ by setting $\sigma_\even(v) = w$ for $v \in V_\even
\cap \winsubeven{G}$ and $w \in \post{v}$ such that $\rho(w) \le \rho(w')$
for all $w' \in \post{v}$.  A winning strategy for player $\odd$
cannot be extracted from $\rho$ \emph{a posteriori}, so, as also
observed in~\cite{Sch:07}, the runtime
complexity of computing a winning strategy cannot be improved by
considering the dual of a game (contrary to what is stated in~\cite{Jur:00}). 
\begin{theorem}[See also~\cite{Sch:07}]
Algorithm~\ref{alg:spm} can compute
winning strategies for both players in~$\runtimeceil$. 
\end{theorem}
As an illustration of the above observations, consider the family of games
depicted in Figure~\ref{fig:tops_faster}. The more optimal runtime
of $\runtimefloor$ will be achieved by solving the games themselves
(using $\mathbb{M}^\even$) and not their dual. As all games in the family are
$\odd$-paradises, we cannot extract a winning strategy for player
$\odd$ from the computed progress measure and the only option we have is to 
solve the dual games with the less favourable runtime of $\runtimeceil$.
In fact, all instances of the family of games depicted in
Figure~\ref{fig:tops_faster} are solved exponentially faster than
their dual, underlining the potential practical implications of re-running
the algorithm.

To facilitate the analysis of SPM, we will use the following  terms and notions. The term \emph{measure} refers to the intermediate values of $\rho$ in the lifting process as well. Given a tuple $m \in \mathbb{M}^\even$, we say that the position $i$ in $m$ is \emph{saturated}, if $(m)_i = |V_i|$.

\section{An operational interpretation of progress measures}
\label{sec:interpretation}

\def\globmaxprio{max \priosym}

\def\eventups{\mathbb{M}^{\even}}
\def\exteventups{\mathbb{M}^{\even}_{ext}}
\def\oddtups{\mathbb{M}^{\odd}}

\def\allplays{\Pi}
\def\profilesym{\theta}
\newcommand{\profilefun}[1]{\profilesym_{#1}}
\newcommand{\profile}[2]{\profilesym_{#1}(#2)}

\newcommand{\maxval}[2]{\varphi^{*}_{#1}(#2)}

\newcommand{\succtup}[2]{\textsf{succ}_{#1}(#2)}

While, from a technical perspective, SPM is a relatively simple
algorithm in the sense that it is
concise and its individual steps are elementary operations, it lacks
a clear and appealing explanation of the devices used. It is therefore
difficult to understand, and possibly enhance. Apart from the formal
definition of progress measures, little explanation
of what is hidden behind the values in tuples is offered.
Notable exceptions are \cite{Klau:01}, which explains that for
$\odd$-solitaire games with only even simple cycles, one can use the 
maximal degrees of `odd
stretches' (a concept we make precise below) in order to define a
parity progress measure, and Schewe's bigstep paper \cite{Sch:07},
where it is shown that dominions of a bounded size can be detected
using measures with a restricted codomain. Klauck's construction
for a specific parity progress measure breaks down for arbitrary parity
games and the constructed parity progress measure is not related to
the measure that is computed by the SPM algorithm, nor to any of the
intermediate measures. In general, the  high-level
intuition is that the larger progress measure values indicate more
capabilities of player $\odd$, and a value at a given position in the
tuple is somehow related to the number of priorities that $\odd$
can enforce to visit.\medskip

In what follows, we present a precise and operational interpretation
of measures. Our interpretation is similar in spirit to the one used in \cite{Klau:01},
but applicable to all parity games, and uses a concept known
from the realm of strategy improvement algorithms -- a value (or
profile) of a play. Here, values are defined in terms of maximal odd-dominated
stretches occurring in a prefix of a play. With this notion at hand,
we can consider an optimal valuation of vertices, being the best
lower bound on play values that player $\even$ can enforce, or the
worst upper bound that $\odd$ can achieve, \ie it is an \emph{equilibrium}.
The optimal valuations range over the same codomain as progress
measures, and the main result of this section states that the least
game parity progress measure is equal to the optimal valuation.\medskip

Let $\exteventups$ denote all tuples in $\nat^{d} \cup \{\top\}$
such that for all $m \in \exteventups$ and even positions $i \leq
d$, $(m)_i =0$; \ie compared to $\eventups$, we drop the requirement that
values on odd positions $i$ are bounded by $|V_i|$. Elements in $\exteventups$ are ordered in the same fashion as those in $\eventups$. Given a 
play $\pi$, a \emph{stretch} is a subsequence
$\pi_s \pi_{s+1} \dots \pi_{s+l}$ of $\pi$. For a priority $k$, a
\emph{$k$-dominated stretch} is a stretch in which the minimal
priority among all vertices in the stretch is $k$. The \emph{degree}
of a $k$-dominated stretch is the number of vertices with priority
$k$ occurring in the stretch.
\begin{definition}
Let us denote all plays in the parity game by $\allplays$. An $\even-$\emph{value} (or simply value) of a play is a function $\profilefun{\even}: \allplays \longrightarrow \exteventups$ defined as follows:
\begin{itemize}
 \item if $\pi$ is winning for $\odd$, then $\profile{\even}{\pi} = \top$
 \item if $\pi$ is winning for $\even$, then $\profile{\even}{\pi} = m$, where 
 $m \neq \top$, and for every odd position $i$, $(m)_i$ is the \emph{degree} of the maximal $i$-dominated stretch that is a prefix of $\pi$ 
\end{itemize}
\end{definition}
Observe that a play value is well-defined, as an
infinite $i$-dominated stretch for an odd $i$ implies that a game
is won by $\odd$, and its value is $\top$ in such case. 

\begin{example}
Suppose that the sequence of priorities corresponding to a certain
play $\pi$ is $453453213(47)^{*}$. Then $\profile{\even}{\pi} =
(0,1,0,2,0,0,0,0)$.

\end{example}

The theorem below links the progress measure values to players' capabilities to enforce beneficial plays or avoid harmful ones, where the benefit from a play is measured by a specially devised play value, as it is done in strategy improvement algorithms. 
This offers a more operational view on progress measure values, which, combined with a more fine-grained analysis of the mechanics of SPM allows us to extract winning strategies for both players in the next section.

\begin{theorem}
\label{thm:progint}
If $\overline{\rho}$ is the least progress measure of a parity game $G$, then, for all $v$:
\begin{enumerate}
 \item there is a strategy $\sigma_{\odd} \in \memstrategy{\odd}$ such that for every $\pi \in \plays{\sigma_{\odd}}{v}$, $\profile{\even}{\pi} \geq \overline{\rho}(v)$
 \item there is a strategy $\sigma_{\even} \in \strategy{\even}$ such that for every $\pi \in \plays{\sigma_{\even}}{v}$, $\profile{\even}{\pi} \leq \overline{\rho}(v)$
\end{enumerate}
\end{theorem}

A notable difference between strategy improvement algorithms and SPM is
that SPM does not work with explicit strategies, and the intermediate
measure values do not represent any proper valuation induced by
strategies -- only the final least progress measure does. Still,
these intermediate values give an underapproximation of the
capabilities of player $\odd$ in terms of odd-dominated stretches
that she can enforce.

Note that a consequence of Theorem \ref{thm:progint} is that the least (resp. greatest) play values that player $\odd$ (resp. $\even$) can enforce are equal, and coincide with the least game parity progress measure $\overline{\rho}$ computed by SPM. 
Observe also that player $\even$ can always achieve the strategy guaranteeing the optimal even-biased play value with a memoryless strategy, whereas player $\odd$ may require to that end a strategy that depends on a play's history.

\section{Strategy construction for player \odd}
\label{sec:strategy}

Computing winning strategies is typically part of a practical
solution to a complex verification or a controller synthesis problem.
In such use cases, these strategies are employed to
construct witnesses and counterexamples for the verification problems,
or for constructing control strategies for the controller~\cite{AVW:03}.
As we explained in Section \ref{sec:spm}, the SPM algorithm can
easily be extended to construct a winning strategy for player
$\even$. The problem of deriving a winning strategy for player \odd
in SPM (other than by running the algorithm on the `dual' game, or
by using a `dual' domain $\mathbb{M}^{\odd}$) has, however, never
been addressed. Note that the problem of computing strategies is
at least as hard as solving a game. Indeed, even if we are equipped
with a valid winning partition $(\winsubeven{G},\winsubodd{G})$ for
a game $G$, then deriving the strategies witnessing these partitions
involves the same computational overhead as the one required to
compute $(\winsubeven{G},\winsubodd{G})$ in the first place.
\begin{restatable}{proposition}{propstrategiesdifficult} 
\label{prop:strategies_difficult}
The problem of finding the winning partition
$(\winsubeven{G},\winsubodd{G})$ of a given game $G$ can be reduced
in polynomial time to the problem of computing a winning strategy
for player $\player$ in a given $\player$-dominion.
\end{restatable}

Deriving a strategy for both players by using the SPM to compute
$\mathbb{M}^{\even}$ measures and $\mathbb{M}^{\odd}$ measures
consecutively, or even simultaneously, affects, as we already
discussed in Section~\ref{sec:spm}, SPM's runtime.    Being able
to compute \odd strategies without resorting to the aforementioned
methods would also allow us to potentially significantly improve
efficiency on such instances as given by Figure~\ref{fig:tops_faster}.
It may be clear, though, that extracting a winning strategy from
the small progress measures algorithm for vertices with measure
$\top$ will require modifying the algorithm (storing additional
data, augmenting the lifting process).  For instance, simply following
the vertex that caused the update to top, fails, as the example
below shows.

\begin{example} \label{ex:greedytop_wrong} Reconsider the game
depicted in Figure~\ref{fig:example}. Recall that 
vertices $v_4, v_5$ and $v_6$ are won by player $\odd$, and in all
possible lifting schemes, the first vertex whose measure becomes
$\top$ is $v_6$. After that, a possible sequence of liftings is
that first $\rho(v_5)$ is set to $\top$ (due to $v_6$), followed
by $\rho(v_4) = \top$ (due to $v_5$). If we set the strategy
based on the vertex that caused the given vertex to be lifted to top,
we obtain $\sigma(v_4) = v_5$, which is not winning for
$\odd$.

\end{example}
The key problem is that for vertices won by player \odd, from some
point onward, the lifting process discards significant information.
This is best seen in case of lifting to $\top$ -- a partial
characterisation of reachable odd priorities contained in a tuple
(see also our previous section) is ultimately replaced with a mere
indication that player \odd can win.

\subsection{A Bounded $\odd$-Dominion}

Consider a game $G$ on which a standard SPM algorithm with an arbitrary lifting policy has been applied. Suppose that at some point a vertex $v$ is the first one to be lifted to $\top$, and after lifting of $v$ the algorithm is suspended, resulting in some temporary measure $\rho$. Let $k$ be the priority of $v$. 

We will start with two straightforward observations. The first one
is that $k$ must be an odd number; this is because a vertex with an
even priority obtains, after lifting, a $\rho$-value equal to the
$\rho$-value of one of its successors, and therefore it cannot be
the first vertex lifted to $\top$. Another immediate conclusion is
that at least one (or all, if $v \in V_{\even}$) successor(s) of
$v$ has (have)  a $\rho$-value saturated up to the $k$-th position,
\ie it is of the form $m = (0,|V_1|, 0, |V_3|,\dots, 0, |V_k|, ***)$;
were it not the case, then a non-top value $m'$ such that $m' >_{k}
m$ would exist, which would be inconsistent with the definition of
\progname.

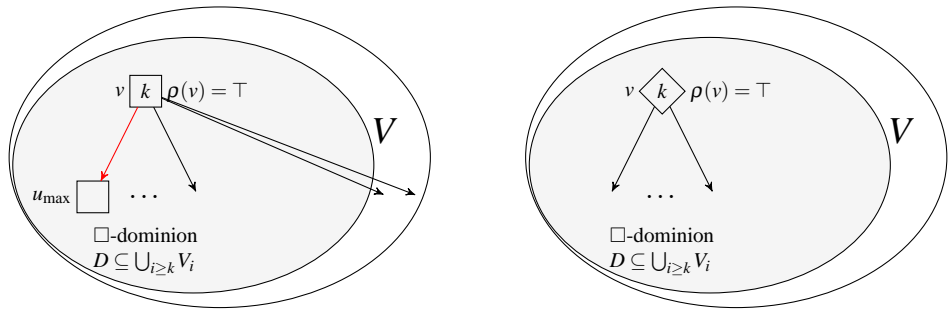
\begin{figure}[h!]
\centering
\begin{tikzpicture}[>=stealth']
\tikzstyle{every node}=[draw, inner sep=2pt, outer sep=0pt, node distance=40pt];

\node[draw=none] (x) {};
\node[draw=none] (w) [below of=x,xshift=-10pt,yshift=37pt] {};
\path[draw,fill=gray!8] (w) ellipse (2.4cm and 1.7cm) ;
\node[shape=rectangle,minimum size=12pt,label=left:{\scriptsize $v$},label=right:{\scriptsize $\rho(v) = \top$}] (y) [above of=x,xshift=-28pt,yshift=-15pt] {\scriptsize $k$};

\node[shape=rectangle,minimum size=12pt,label=left:{\scriptsize $u_{\text{max}}$}] (z1) [below of= y,xshift=-20pt] {};

\node[draw=none] (z2) [below of=y,xshift=0pt] {$\dots$};
\node[draw=none] (z3) [below of=y,xshift=20pt] {};

\node[draw=none] (z5) [below of=y,xshift=92pt] {};
\node[draw=none] (z6) [below of=y,xshift=104pt] {};
\node[draw=none] (u) [below of=z2,yshift=20pt] {\scriptsize \begin{tabular}{l}$\odd$-dominion \\ $D \subseteq \bigcup_{i \ge k} V_i$ \end{tabular} };

\path[draw] (x) node[draw=none,xshift=63pt,yshift=10pt] {\Large $V$} ellipse (2.8cm and 2.0 cm);

\draw[->] (y) edge[color=red] (z1) edge  (z3) edge (z5) edge (z6)
;

\end{tikzpicture}
\qquad\quad
\begin{tikzpicture}[>=stealth']
\tikzstyle{every node}=[draw, inner sep=2pt, outer sep=0pt, node distance=40pt];

\node[draw=none] (x) {};
\node[draw=none] (w) [below of=x,xshift=-10pt,yshift=37pt] {};
\path[draw,fill=gray!8] (w) ellipse (2.4cm and 1.7cm) ;
\node[shape=diamond,minimum size=17pt,label=left:{\scriptsize $v$},label=right:{\scriptsize $\rho(v) = \top$}] (y) [above of=x,xshift=-28pt,yshift=-15pt] {\scriptsize $k$};
\node[draw=none] (z1) [below of=y,xshift=-20pt] {};
\node[draw=none] (z2) [below of=y,xshift=0pt] {$\dots$};
\node[draw=none] (z3) [below of=y,xshift=20pt] {};
\node[draw=none] (u) [below of=z2,yshift=20pt] {\scriptsize \begin{tabular}{l}$\odd$-dominion \\ $D \subseteq \bigcup_{i \ge k} V_i$ \end{tabular} };

\path[draw] (x) node[draw=none,xshift=63pt,yshift=10pt] {\Large $V$} ellipse (2.8cm and 2.0 cm);

\draw[->] (y) edge (z1) edge  (z3)
;

\end{tikzpicture}
\caption{Snapshot of the SPM algorithm after the first vertex $v$ is lifted to top.}
\label{fig:egg}
\end{figure}

There are two more complex properties, which we can utilise to modify the SPM algorithm and compute the winning strategy for player \odd (see Figure \ref{fig:egg}). 
\begin{enumerate}
 \item Vertex $v$ belongs to an \odd-dominion $D$ within $G$ such that the minimal priority in $D$ is $k$.
 \item If $v \in V_{\odd}$, then picking the successor $u_{max}$ of $v$ with the maximal current $\rho$-value among $\post{v}$ is a part of a (positional) winning strategy for \odd that stays within such a dominion $D$ as described above. 
 \end{enumerate}
The intuition concerning the above facts is as follows. Vertices with
a measure value $m$ saturated up to but possibly excluding a certain
position $i$ have a neat interpretation of the measure value at
position $i$: \medskip

\noindent\emph{Player \odd can force the following outcome of a play:
\begin{enumerate}
 \item priority $i$ appears $m_i$ times without any lower priority in between
 \item the play will reach a $\top$-labelled vertex via priorities not more significant than $i$
 \item the play enters a cycle with an odd dominating priority larger (less significant) than $i$.
\end{enumerate}
}\medskip

\begin{figure}[h!]
\centering
\begin{tikzpicture}[>=stealth']
\tikzstyle{every node}=[draw, inner sep=1pt, outer sep=1pt, node distance=40pt];
  \node[shape=diamond,minimum size=26,label=above:{$v_1$},label=below:{\scriptsize $\top$}] (y1)                            {\scriptsize 1};
  \node[shape=rectangle,minimum size=22,label=above:{$v_2$},label=below:{\scriptsize $(0,2,0,0)$~~~}] (y2) [right of=y1,xshift=20pt]  {\scriptsize 2};
  \node[shape=diamond,minimum size=26,label=below:{$v_3$},label=right:{\scriptsize $(0,2,0,0)$}] (y3) [above right of=y2,xshift=35pt]  {\scriptsize 1};
  \node[shape=diamond,minimum size=26,label=above:{$v_4$},label=right:{\scriptsize $(0,2,0,1)$}] (y4) [below right of=y2,xshift=35pt]  {\scriptsize 3};

\path[->]
  (y1) edge[color=red] (y2) 
  (y2) edge (y3) edge[bend left,color=red] (y4)
  (y3) edge[bend right] (y1)
  (y4) edge[bend left] (y2)
;
\end{tikzpicture}
\caption{A game, won entirely by player $\odd$, and demonstrating
that a strategy defined by a greedy choice of vertex with the maximal 
tuple does not work. After lifting the vertices in order $v_1,v_3,v_2,v_4,v_1$, we obtain the measure values as above. Player \odd would then choose $v_3$, which leads to a losing play, whereas the choice of the other successor $(v_4)$ yields a winning play for \odd.}
\label{fig:greedywrong}
\end{figure}
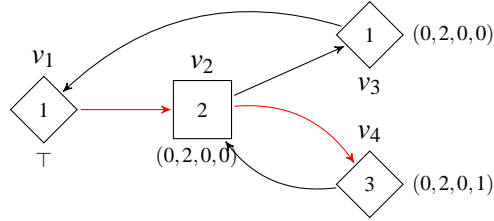

\noindent
Therefore, in the situation as described above, \odd can force a
play starting at $v$ to first go in one step to the successor
$u_{max}$ of $v$ with a measure of the form $(0,|V_1|, 0, |V_3|,
\dots, 0, |V_k|, ***)$, and then to play further and either force
a less significant odd-dominated cycle (cases 2 and 3, since $v$
is the only $\top$-labelled vertex), or to visit vertices with
priority $k$ $|V_k|$ times without any lower priority in between.
But in the latter case, since $v$ has priority $k$, we have in fact
$|V_k| + 1$ vertices with priority $k$ not ``cancelled'' by a lower
priority. Hence player \odd has forced an odd-dominated cycle with
the lowest (most significant) priority $k$.  Note that this does
not imply we can simply construct a winning strategy for $\odd$ by
always picking a successor with the maximal measure to further
vertices that can be visited along the play; such a method may lead
to an erroneous strategy, as illustrated by Figure \ref{fig:greedywrong}.

Propagating a top value only to vertices with less significant
priorities is, however, safe. This can be achieved efficiently by
a slightly modified attractor that works within a given context of vertices $W$, 
which we call a \emph{guarded} attractor.
\begin{definition}
Let $k$ be some priority and $U,W$ some sets for which
$U \subseteq W \cap V_{\geq k}$. 
Then $\myattrW{\ge k}{\odd}{U}{W}$ is the least set $A$ satisfying:
\begin{enumerate}
 \item $ U \subseteq A \subseteq W \cap V_{\geq k}$
 
 \item 
 \begin{enumerate}
  \item if $u \in V_{\odd}$ and $\post{u} \cap A \neq \emptyset$, then $u \in A$
  \item if $u \in V_{\even}$ and $\post{u}\cap W \subseteq A$, then $u \in A$
 \end{enumerate}
\end{enumerate}
If $W = V$, we drop the subscript $W$ from the guarded attractor.
\end{definition}

\newcommand{\exttuples}[1]{\mathbb{M}^{#1,\textsf{ext}}}
\def\extprog{newProg}

\noindent
The theorem below forms the basis of our algorithm; it describes
the relevant information about an \odd-dominion that can be extracted
once the first vertex in the game is lifted to top.

\begin{restatable}{theorem}{thmcentraal}
 \label{thm:centraal}
  Let $G$ be a parity game on which an arbitrary lifting sequence is applied, such that at some point a vertex $v$ with $\prio{v}=k$ is the first vertex whose measure value becomes top. Let $\rho$ be the temporary measure at that point. The following holds:
  
  \begin{itemize}
   \item if $v \in V_{\odd}$, then for every successor $u$ of $v$ with a maximal measure among $\post{v}$ there is an \odd-dominion $D_u$ containing $\myattr{\geq k}{\odd}{\{v\}}$ such that for all $w \in D_u$, $\prio{w} \geq k$. Moreover, $\odd$ has a winning strategy that is closed on $D_u$, and which is defined on $v$ as $\sigma(v) = u$, and on $\myattr{\geq k}{\odd}{\{v\}}  \setminus \{v\}$ as the strategy attracting towards $v$,
   \item if $v \in V_{\even}$, then there is an \odd-dominion $D$ containing $\myattr{\geq k}{\odd}{\{v\}}$ such that for all $w \in D$, $\prio{w} \geq k$. Moreover,  \odd has a winning strategy $\sigma$ that is closed on $D$, and defined on $\myattr{\geq k}{\odd}{\{v\}}  \setminus \{v\}$ as the strategy attracting towards $v$.  Note that in this case $\post{v} \subseteq D$. 
   \end{itemize}     
     
 \end{restatable} 

\subsection{The Extended SPM Algorithm}

\def\res{\textsf{RES}}
\def\rem{\textsf{REM}}
\def\irr{\textsf{IRR}}
\def\dom{\textsf{DOM}}
\def\newspm{\textsf{SPM-Within}}

\begin{algorithm}[h!t]
\caption{Modified SPM with strategy derivation for player Odd}
\label{alg:newspm}
\begin{algorithmic}[1]
\Function{Solve}{$G$}
\State \emph{\textbf{Input} $G = (V, E, \priosym, (V_\even, V_\odd))$}
\State \emph{\textbf{Output} Winning partition and strategies $((\winsubeven{G},\sigma'),(\winsubodd{G},\sigma))$}
\State initialise $\sigma$ to an empty assignment
\State $\rho  \gets \lambda w \in V.~(0, \dots, 0)$
\State $\Call{\newspm}{V}$
\State compute strategy $\sigma'$ for player Even by picking min. successor w.r.t. $\rho$
\State \Return $((\{v \in V ~|~ \rho(v) \neq \top\},\sigma'),(\{v \in V ~|~ \rho(v) = \top,\sigma))$

\State

\Procedure{\newspm}{$W$}  

\While{$(W \neq \emptyset)$} \label{line:outerwhile}
\While{$\rho \sqsubset \lift{\rho}{w} \text{ for some $w \in W$ and for all $w \in W$:} \rho(w) \neq \top$} \label{line:liftloop_start}
\State $\rho \gets \lift{\rho}{w}$ for $w \in W$ such that $\rho \sqsubset \lift{\rho}{w}$
\EndWhile \label{line:liftloop_end}
\State \textbf{if} {$\text{for all $w \in W$: } \rho(w) \neq \top$} \textbf{break} \textbf{end if}

\State \label{line:firsttop_v} $v \gets $ the (unique) vertex in $W$ such that $\rho(v) = \top$
\State $k \gets \prio{v}$
\State \label{line:firsttop_strat} $\sigma(v) \gets u$ where $u \in v^\bullet \cap W$ for which $\rho(u') \leq_{k} \rho(u)$ for all $u' \in v^\bullet \cap W$
\State $\res \gets \myattrW{\geq k}{\odd}{\{v\}}{W}$
\For{ \textbf{all} $w \in \res \setminus \{v\}$}
\State $\rho(w) \gets \top$
\State \textbf{if} $w \in V_{\odd}$ \textbf{then} assign $\sigma(w)$ the strategy \emph{attracting to $v$} \textbf{end if} 
\EndFor \label{line:defsigma_res_end}
\State $\dom \gets \res$
  \State \label{line:compute_irr} $\irr \gets \attrW{\even}{\{ w \in W \,\mid\, \prio{w} < k\}}{W}$
  \State \label{line:rem} $\rem \gets W \setminus (\res \cup \irr)$
  \State \label{line:reccall} $\Call{\newspm}{\rem}$  
  \State $\dom \gets \dom \cup \{ w \in \rem \,\mid\, \rho(w) = \top \}$ \label{line:dominion_resolved}
\State $A \gets \attrW{\odd}{\dom}{W}$

  \For{ \textbf{all} $w \in A \setminus \dom$}
  \State $\rho(w) \gets \top$
    \If{$w \in V_{\odd}$} 
    assign $\sigma(w)$ to be the strategy  \emph{attracting} to $\dom$ 
    \EndIf
  \EndFor
\State \label{line:smallerW} $W \gets W \setminus A$
\EndWhile
\EndProcedure
\EndFunction

\end{algorithmic}
\end{algorithm}

\newcommand{\genlift}[3]{\ensuremath{\text{lift}_{#1}(#2,#3)}}
\def\genliftname{\text{lift}_{W}}

Theorem \ref{thm:centraal} captures the core idea behind our algorithm.
It provides us with the means to locally resolve (\ie define a local
strategy for) at least one vertex in $\winsubodd{G}$, once a top value
is found while lifting. Moreover, it indicates in which part of
the game the remainder of the \odd-dominion resides, implying
that one can temporarily restrict the lifting to that area until
the dominion is fully resolved. 
We will give a description of our solution (Algorithm \ref{alg:newspm}), and
informally argue the correctness of our approach.  For a (intricate and rather involved) formal proof, we
refer to~\cite{GW:14}.

The algorithm proceeds as follows. First, a standard SPM is run until the first vertex reaches top [l. \ref{line:liftloop_start}--\ref{line:liftloop_end} in Alg. \ref{alg:newspm}\,]. Whenever $v$ is the first vertex lifted to top, then the issue of the winning strategy for $v$ can be resolved immediately [l. \ref{line:firsttop_strat}\,], as well as for vertices in the guarded attractor of $v$ (if there are any). We will denote this set of `resolved' vertices with $\res$. Moreover, we can restrict our search for the remainder of the \odd-dominion $D$ only to vertices with priorities not more significant than $k$, in fact only those from which player \even cannot attract a play to visit a priority more significant than $k$. Hence we can remove from the current computation context the set $\irr = \myattr{}{\even}{\{ w \in W \,\mid\, \prio{w} < k\}}$, vertices that may be considered at the moment irrelevant [l. \ref{line:compute_irr}--\ref{line:rem}\,].

After discarding the resolved and currently irrelevant vertices, the algorithm proceeds in the remaining set of vertices that constitutes a proper subgame (i.e. without dead ends) induced by the set $\rem$. After the subroutine returns [l. \ref{line:reccall}\,], all vertices labelled with top are won by player \odd in the subgame $G \cap \rem$. In other words, those vertices are won by \odd provided that the play does not leave $\rem$. Since the only way for player \even to escape from $\rem$ is to visit $\res$, where every vertex is won by player \odd, the top-labelled vertices from $\rem$ are in fact won by \odd in the context of the larger game $G \cap W$. Therefore the set $\dom$ computed in line \ref{line:dominion_resolved} constitutes an \odd-dominion within the game $G \cap W$, in which we have moreover fully defined a winning strategy $\sigma$ for player \odd. Finally, every vertex from $V \setminus \dom$ that can be attracted by player \odd to the dominion $\dom$ is certainly won by \odd, and for those vertices we assign the standard strategy attracting to $\dom$. The \odd-dominion $A$ is then removed, and the computation continues in the remaining subgame. 
\medskip

The algorithm may at first sight appear to deviate much from the
standard SPM algorithm. However, the additional overlay, apart from
defining the strategy, is no more than a special lifting policy that
temporarily restricts the lifting to parts where an \odd dominion
resides. \medskip

Every attractor computation takes $O(n+m)$
time, and whenever it occurs, at least one new vertex is `resolved'.
Hence the total extra time introduced by the attractor computations
is bounded by $O(n(n+m))$. As with the standard SPM, the lifting
operations dominate the running time, and their total number for
every vertex is bounded by the size of $\mathbb{M}^{\even}$.
\begin{theorem}
For a game $G$ with $n$ vertices, $m$ edges, and
$d$ priorities, $\textsc{Solve}$ solves $G$ and computes winning
strategies for player $\even$ and $\odd$ in worst-case $\runtimefloor$.
\end{theorem}

\section{Illustrating the new algorithm}
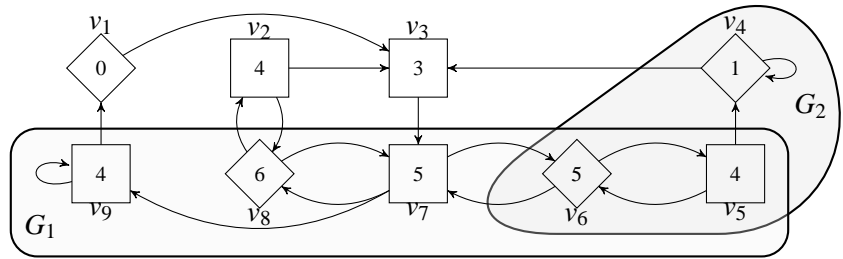
\begin{figure}[h!tbp]
\centering
\begin{tikzpicture}[>=stealth']
\tikzstyle{every node}=[draw, inner sep=0pt, outer sep=0pt, node distance=40pt];
  \node[draw=none,shape=diamond,minimum size=26]   (s1)                                 {};
  \node[draw=none,shape=rectangle,minimum size=22] (s2) [right of=s1,xshift=20pt]                   {};
  \node[draw=none,shape=rectangle,minimum size=22] (s3) [right of=s2,xshift=20pt]                   {};
  \node[draw=none,shape=diamond,minimum size=26]   (s4) [right of=s3,xshift=80pt]                   {};
  \node[draw=none,shape=rectangle,minimum size=22] (s5) [below of=s4]                   {};
  \node[draw=none,shape=diamond,minimum size=26]   (s6) [left of=s5,xshift=-20pt]                   {\scriptsize 5};
  \node[draw=none,shape=rectangle,minimum size=22] (s7) [left of=s6,xshift=-20pt]                   {};
  \node[draw=none,shape=diamond,minimum size=26]   (s8) [left of=s7,xshift=-20pt]                   {};
  \node[draw=none,shape=rectangle,minimum size=22] (s9) [left of=s8,xshift=-20pt]                   {};

\draw [fill=gray!15, thick,rounded corners=55pt, fill opacity=0.5] 
    ($(s5)+(1.7,-0.8)$) -- 
    ($(s4)+(1.0,1.65)$) -- 
    ($(s6)+(-2.2,-0.8)$) -- cycle
;
\draw 
    [fill=gray!10, thick,rounded corners=10pt, fill opacity=0.3] 
    ($(s5)+(0.7,-1.1)$) -- 
    ($(s9)+(-1.2,-1.1)$) -- 
    ($(s9)+(-1.2,0.6)$) -- 
    ($(s5)+(0.7,0.6)$) -- cycle;

\draw ($(s4)+(1.0,-0.5)$) node[draw=none] {$G_2$};
\draw ($(s9)+(-0.8,-0.7)$) node[draw=none] {$G_1$};

  \node[label=above:{$v_1$},shape=diamond,minimum size=26]   (y1)                                 {\scriptsize 0};
  \node[label=above:{$v_2$},shape=rectangle,minimum size=22] (y2) [right of=y1,xshift=20pt]                   {\scriptsize 4};
  \node[label=above:{$v_3$},shape=rectangle,minimum size=22] (y3) [right of=y2,xshift=20pt]                   {\scriptsize 3};
  \node[label=above:{$v_4$},shape=diamond,minimum size=26]   (y4) [right of=y3,xshift=80pt]                   {\scriptsize 1};
  \node[label=below:{$v_5$},shape=rectangle,minimum size=22] (y5) [below of=y4]                   {\scriptsize 4};
  \node[label=below:{$v_6$},shape=diamond,minimum size=26]   (y6) [left of=y5,xshift=-20pt]                   {\scriptsize 5};
  \node[label=below:{$v_7$},shape=rectangle,minimum size=22] (y7) [left of=y6,xshift=-20pt]                   {\scriptsize 5};
  \node[label=below:{$v_8$},shape=diamond,minimum size=26]   (y8) [left of=y7,xshift=-20pt]                   {\scriptsize 6};
  \node[label=below:{$v_9$},shape=rectangle,minimum size=22] (y9) [left of=y8,xshift=-20pt]                   {\scriptsize 4};

\path[->]
  (y1) edge[bend left] (y3)
  (y2) edge (y3) edge[bend left] (y8)
  (y3) edge (y7)
  (y4) edge (y3) edge[loop right] (y4)
  (y5) edge (y4) edge[bend left] (y6)
  (y6) edge[bend left] (y5) edge[bend left] (y7)
  (y7) edge [bend left] (y6) edge[bend left] (y8) edge[bend left] (y9)
  (y8) edge [bend left] (y2) edge[bend left] (y7)
  (y9) edge (y1) edge[loop left] (y9)
;

\end{tikzpicture}
\caption{An example game $G$ with two (overlapping) subgames $G_1$ and $G_2$.}
\label{fig:illustrating_example}
\end{figure}

We illustrate the various aspects of Algorithm~\ref{alg:newspm} on
the game $G$ depicted in Figure~\ref{fig:illustrating_example}, with two (overlapping) subgames
$G_1$ and $G_2$.
Note that the entire game is an $\odd$-paradise: every vertex
eventually is assigned measure $\top$ by Algorithm~\ref{alg:newspm}
(and Algorithm~\ref{alg:spm}, for that matter). Suppose we use a
lifting strategy prioritising $v_2,v_3,v_7$ and $v_8$; then vertex
$v_3$'s measure is the first to reach $\top$, and the successor
with maximal measure is $v_7$. Therefore, $\odd$'s strategy is to
move from $v_3$ to $v_7$.  The set $\res$, computed next consists of
vertices $v_3$ and $v_2$; the strategy for $v_2$ is set to $v_3$
and its measure is set to $\top$. The $\even$-attractor into those
vertices with priorities $\ge 3$, \ie, vertices
$v_1$ and $v_4$, is exactly those vertices, so, next, the algorithm
zooms in on solving the subgame $G_1$.

Suppose that within the latter subgame, we prioritise the lifting of
vertex $v_7$ and $v_8$; then vertex $v_7$'s measure is set to $\top$
first, and $v_7$'s successor with the largest measure is $v_8$;
therefore $\odd$'s strategy is to move from $v_7$ to $v_8$. At this
point in the algorithm, $\res$ is assigned the set of vertices $v_7$
and $v_8$, and the measure of $v_8$ is set to $\top$. Note that in this
case, in this subgame, the winning strategy for $\odd$ on $v_7$ is to remain within
the set $\res$. Since all
remaining vertices have more signficant priorities than $v_7$, or
are forced by $\even$ to move there, the next recursion is run on
an empty subgame and immediately returns without changing the
measures. Upon return, the $\odd$-attractor to all $\odd$-won
vertices (within the subgame $G_1$, so these are only the 
vertices $v_7$ and $v_8$)
is computed, and the algorithm continues solving the remaining
subgame (\ie the game restricted to vertices $v_5,v_6$ and $v_9$),
concluding that no vertex in this entire game will be assigned
measure $\top$.

At this point, the algorithm returns to the global game again and
computes the $\odd$-attractor to the vertices won by player $\odd$ at that stage
(\ie vertices $v_2,v_3,v_7$ and $v_8$), adding vertices $v_1$ and $v_9$,
setting their measure to $\top$ 
and setting $\odd$'s strategy for $v_9$ to move to $v_1$. 

As a final step, the algorithm next homes in on the subgame $G_2$,
again within the larger game.  The only vertex assigned measure
$\top$ in the above subgame is vertex $v_4$; at this point $\res$ is
assigned all vertices in $G_2$, the measure of $v_5$ and $v_6$ is set to
$\top$ and the $\odd$ strategy for vertex $v_5$ is set to $v_4$.
This effectively solves the entire game.

\section{Conclusions and Future Work}
In this paper, we studied the classical Small Progress Measures algorithm for solving
parity games.  The two key contributions of our work are as follows:
\begin{enumerate}
 \item We have proposed a more operational interpretation of progress measures by characterising the types of plays that players can enforce. 
 \item We have provided a modification of the SPM algorithm that allows to compute the winning strategies for both players in one pass, thus improving the worst-case running time of strategy derivation.
 \end{enumerate}
The second enhancement has been made possible due to a thorough study of the contents of progress measures, and their underapproximations in the intermediate stages of the algorithm (building on the proposed operational interpretation). \medskip

As for the future work, we would like to perform an analysis of
SPM's behaviour on special classes of games, along the same lines
as we have done in case of the recursive algorithm \cite{GW:13}.
More specifically, we would like to identify the games for which
SPM runs in polynomial time, and study enhancements that allow to
solve more types of games efficiently. It would also be interesting
to see whether the ideas behind our modification of the SPM algorithm
carry over to the algorithm for small energy progress
measures~\cite{BCDGR:11} for mean payoff games.

\bibliographystyle{eptcs}

\newpage

\appendix

\end{document}